# Doping Concentration Modulation in Vanadium Doped Monolayer Molybdenum Disulfide for Synaptic Transistors

*Jingyun Zou[1,#], Zhengyang Cai[1,#], Yongjue Lai[1,#], Junyang Tan[1], Rongjie Zhang[1], Simin Feng[1], Gang Wang[2], Junhao Lin[2], Bilu Liu[1,*], Hui-Ming Cheng[1,3,*]*

Dr. J. Zou, Z. Cai, Y. Lai, J. Tan, R. Zhang, Dr. S. Feng, Prof. B. Liu, Prof. H.-M. Cheng

Shenzhen Geim Graphene Center, Tsinghua–Berkeley Shenzhen Institute & Tsinghua Shenzhen International Graduate School, Tsinghua University, Shenzhen 518055, P. R. China

E-mail: bilu.liu@sz.tsinghua.edu.cn; hmcheng@sz.tsinghua.edu.cn

G. Wang, Prof. J. Lin

Department of Physics, SUSTech Core Research Facilities

Southern University of Science and Technology, Shenzhen 518055, P. R. China

Prof. H.-M. Cheng

Shenyang National Laboratory for Materials Science, Institute of Metal Research, Chinese Academy of Sciences, Shenyang 110016, P. R. China

**Abstract**

Doping is an effective way to modify the electronic property of two-dimensional (2D) materials and endow them with new functionalities. However, wide-range control of the substitutional doping concentration with large scale uniformity remains challenging in 2D materials. Here we report *in-situ* chemical vapor deposition growth of vanadium (V) doped monolayer molybdenum disulfide ($MoS_2$) with widely tunable doping concentrations ranging from 0.3 to 13.1 at%. The key to regulate the doping concentration lies in the use of appropriate V precursors with different doping abilities, which also generate a large-scale uniform doping effect. Artificial synaptic transistors were fabricated by using the heavily doped $MoS_2$ as the channel material for the first time. Synaptic potentiation, depression and repetitive learning processes are mimicked by the gate-tunable channel conductance change in such transistors with abundant V atoms to trap/detrap electrons. This work shows a feasible method to dope monolayer 2D semiconductors and demonstrates their use in artificial synaptic transistors.

## 1. Introduction

Atomically thin two-dimensional (2D) transition metal chalcogenides (TMDCs) have attracted a lot of attention in the past decade. Their unique physical and chemical properties make them promising candidates for next-generation electronic devices.[1-2] Nowadays, great progress has been made in the growth of TMDCs through the chemical vapor deposition (CVD) method.[3-5] However, to exert the utmost performance of 2D TMDCs, modification of their physical properties is required, and doping appears to be a powerful method to tailor the properties of materials and performance of devices.[6-8] Up to date, different doping strategies, e.g., intercalation[9], surface charge transfer doping[10] and substitutional doping[11-12] have been reported, among which substitutional doping is the most stable and non-volatile method[13] and can extend functionalities of TMDCs by incorporation of different elements.[11] For example, substituted niobium (Nb) atoms in $MoS_2$ generate a p-type carrier behavior, and incorporation of vanadium (V) and iron (Fe) atoms endows $MoS_2$ and $WS_2$ with room-temperature ferromagnetism.[14-15] It is clear that doping will extend the application prospects of 2D TMDCs.

Doping concentration is important in the modification of the physical properties of 2D TMDCs. For example, Nb-doped $MoS_2$ with a doping concentration of 2 at% shows an n-type transport behavior[16] while it turns to a p-type behavior when the doping concentration increases to 4.7 at%[17]. Therefore, many efforts have been devoted to modulate the doping concentrations of CVD-grown TMDCs by changing the additive amounts of doping precursors, but high-concentration doping of $MoS_2$ remains a challenge. For example, by increasing the ratio of V and Mo atoms to up to 8:2 in the precursors, the doping concentration is still hard to reach 10 at%.[18] Besides, in this precursor-amount-dependent strategy, with the evaporation of the doping precursor, the supply quantity of doping atoms decreases during the CVD growth process. This causes an inhomogeneous distribution of doping atoms as reported in monolayer Nb-doped $WS_2$ with dopant inhomogeneity from center to edge, resulting in a nonuniform physical property.[19] Therefore, a feasible doping strategy is demanded to realize a high-concentration and uniform doping of 2D TMDCs. Moreover, previous works mainly focus on the modification of the carrier type and carrier concentration in TMDCs, while more efforts are needed in exploring new functionalities of substitutionally doped TMDCs.

In this work, we developed an effective method to achieve the controlled substitutional doping of V atoms in monolayer $MoS_2$. Vanadium precursors with different doping abilities were chosen to control the doping concentration in the V-doped $MoS_2$ (V-$MoS_2$). Lightly doped V-$MoS_2$ with a doping concentration of 0.3 at% was obtained using $V_2O_5$ as the precursor, while heavily doped monolayer V-$MoS_2$ flakes with the doping concentrations ranging from 8.1 to 13.1 at% were grown by using $NH_4VO_3$ or $VCl_3$ as the precursors. With adequate supply of doping precursors, the substituted V atoms are uniformly distributed in the heavily doped V-$MoS_2$ and generate a homogeneous electronic performance, making V-$MoS_2$ an appropriate material for electronic devices. Gate-tunable synaptic transistors were fabricated by using the heavily doped V-$MoS_2$ as the channel material. With electrons being trapped/detrapped by abundant trap centers, the channel conductance is modulated by the gate voltage pulses to simulate the potentiation and depression behaviors of synapses. This work provides a feasible strategy to realize the controlled doping of $MoS_2$ and offers a new platform to design artificial synaptic devices.

## 2. Results and Discussions

A scheme of the *in-situ* CVD growth of V-doped $MoS_2$ is illustrated in **Figure 1**a. $MoO_3$ and S powders were used as the Mo and S precursors, and three different V precursors including $V_2O_5$, $NH_4VO_3$ and $VCl_3$, were chosen as the doping additives. The feature of this strategy is to utilize the different doping abilities of these three V precursors to tune the doping concentration in $MoS_2$. As shown in Figure 1b, $V_2O_5$ has a low doping ability due to its high melting point (690

°C) and the relatively low reaction activity reflected by the high V-O bond energy (627 kJ/mol). Consequently, $V_2O_5$ can realize the low-concentration doping of $MoS_2$ because it evaporates and decomposes slightly at the growth temperature to give a trace amount of V atoms. On the contrary, $NH_4VO_3$ and $VCl_3$ have high doping abilities owing to their relatively low decomposition temperature (200 °C and 350 °C)[20-21], which can provide abundant V atoms to achieve high-concentration doping of $MoS_2$. Moreover, $VCl_3$ decomposes at the growth temperature to produce abundant chemically active gaseous $VCl_4$ (V-Cl bond energy 447 kJ/mol)[22], making it have the highest doping ability among the three precursors. With the addition of these V precursors, monolayer V-doped $MoS_2$ was successfully grown on silicon substrates. Typical triangle V-doped monolayer $MoS_2$ flakes with lateral sizes of ~50 µm, ~25 µm and ~55 µm are observed when using $V_2O_5$, $NH_4VO_3$ and $VCl_3$ as V precursors (Figure 1a inset and Figure S1, Supporting Information), and they are labeled as $V_2O_5$-$MoS_2$, $NH_4VO_3$-$MoS_2$ and $VCl_3$-$MoS_2$ below for simplicity. These results confirm the feasibility of this precursor-dependent strategy to grow V-doped $MoS_2$.

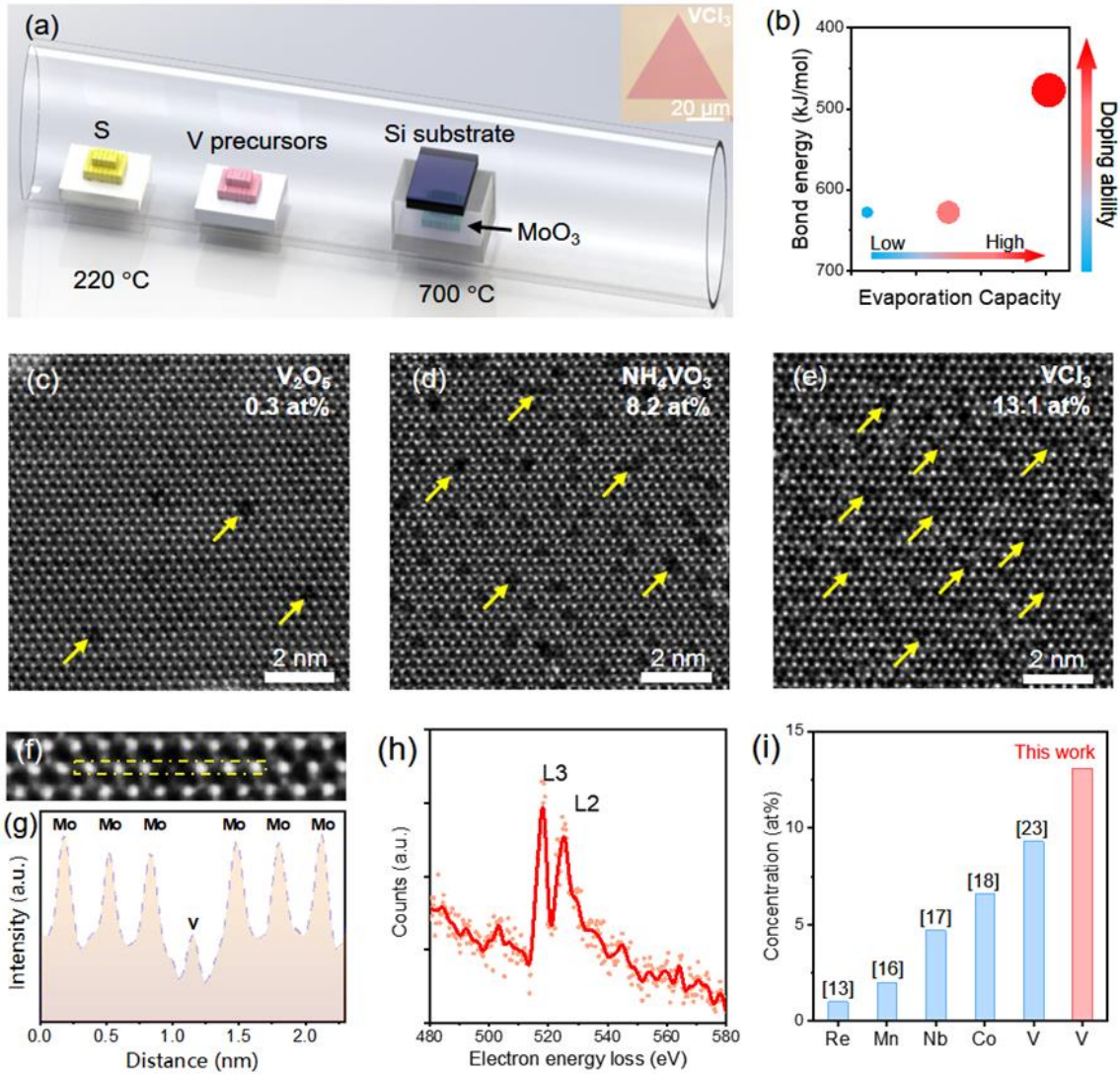

**Figure 1**. Controlled growth of V-doped MoS₂ with different doping concentrations. (a) Illustration of the growth process. The inset shows a typical triangle V-MoS$_2$ grown by using VCl$_3$ as the V precursor. (b) The doping abilities of different V precursors. (c-e) HAADF-STEM images of V-doped MoS$_2$ grown by using V$_2$O$_5$, NH$_4$VO$_3$ and VCl$_3$ as V precursors. Some V atoms with lower intensity are marked by the yellow arrows, and the measured doping concentrations are 0.3 at%, 8.2 at% and 13.1 at%. (f) A magnified STEM image of a typical area with one V atom and (g) the corresponding Z-contrast intensity profile. (h) EELS spectrum of the L2 and L3 edges of V in VCl$_3$-MoS$_2$. (i) Comparison of the doping concentrations of monolayer MoS$_2$ in this work and literature[13, 16-18, 23].

Next, we focus on the doping concentration of V atoms in different $MoS_2$ flakes (Figure 1c-f). V atoms can be directly observed in the high-angle annular dark-field scanning transmission electron microscopy (HAADF-STEM) images by their dimmer contrast (Figure 1f, g), due to the image contrast is related to the atomic number of the imaged species. Their identity in the $MoS_2$ lattice is further validated by the characteristic L-edges of V in the electron energy loss spectrum (EELS) (Figure 1h). With the use of $V_2O_5$ as the doping precursor, only few V atoms are distinguished in the lattice (Figure 1c), resulting in a low doping concentration of 0.3 at%. When changing $NH_4VO_3$ as the V precursor, many V atoms can be identified and the measured doping concentration increases to 8.2 at% (Figure 1d). More V atoms are observed and the doping concentration reaches 13.1 at% when using $NH_4VO_3$ as the precursor (Figure 1e and more details in Figure S2, Supporting Information). To our best knowledge, this value is the highest substitutional doping concentration in semiconducting $MoS_2$ monolayer achieved so far (Figure 1i and Table S1, Supporting Information). These results confirm the efficacy of the appropriate use of precursors to control the doping concentrations of 2D $MoS_2$ in a wide range.

Then we studied how the substituted V atoms affect the bonding state of $MoS_2$. X-ray photoelectron spectroscopy (XPS) data of the pristine $MoS_2$ and the heavily doped $VCl_3$-$MoS_2$ were collected. Signals of the substituted V atoms are observed at 517.0 eV ($2p_{3/2}$) and 524.5 eV ($2p_{1/2}$) in $VCl_3$-$MoS_2$ (**Figure 2**a), further confirming the existence of V atoms. Moreover, for the heavily doped $VCl_3$-$MoS_2$, the signals of Mo 3d and S 2p orbits split compared to the pristine $MoS_2$. Signals of 2H-phase $MoS_2$ (Mo $3d_{5/2}$ at 229.7 eV, Mo $3d_{3/2}$ at 232.9 eV, and S $2p_{3/2}$ at 162.5 eV, $2p_{1/2}$ at 163.6 eV) and some new signals (Mo $3d_{5/2}$ at 229.1 eV, Mo $3d_{3/2}$ at 232.3 eV, and S $2p_{3/2}$ at 161.7 eV, S $2p_{1/2}$ at 162.7 eV) are observed in Figure 2c and d, suggesting that the bonding states of part Mo and S atoms have been changed by the nearby substituted V atoms. Similar phenomena have been observed in Re-doped $MoS_2$ as the coordination of Mo changed from trigonal prismatic to octahedral geometry[24] and cerium doped $BiFeO_3$ due to the formation of oxygen octahedron[25]. After being annealed at 500 °C for 30 min, the XPS spectra of the $VCl_3$-$MoS_2$ (Figure 2a, c and d) and the ratio of the Mo atoms whose bonding states have been affected by V atoms keep unchanged (Figure 2b). The results suggest that substituted V atoms generate a stable doping effect in $MoS_2$.

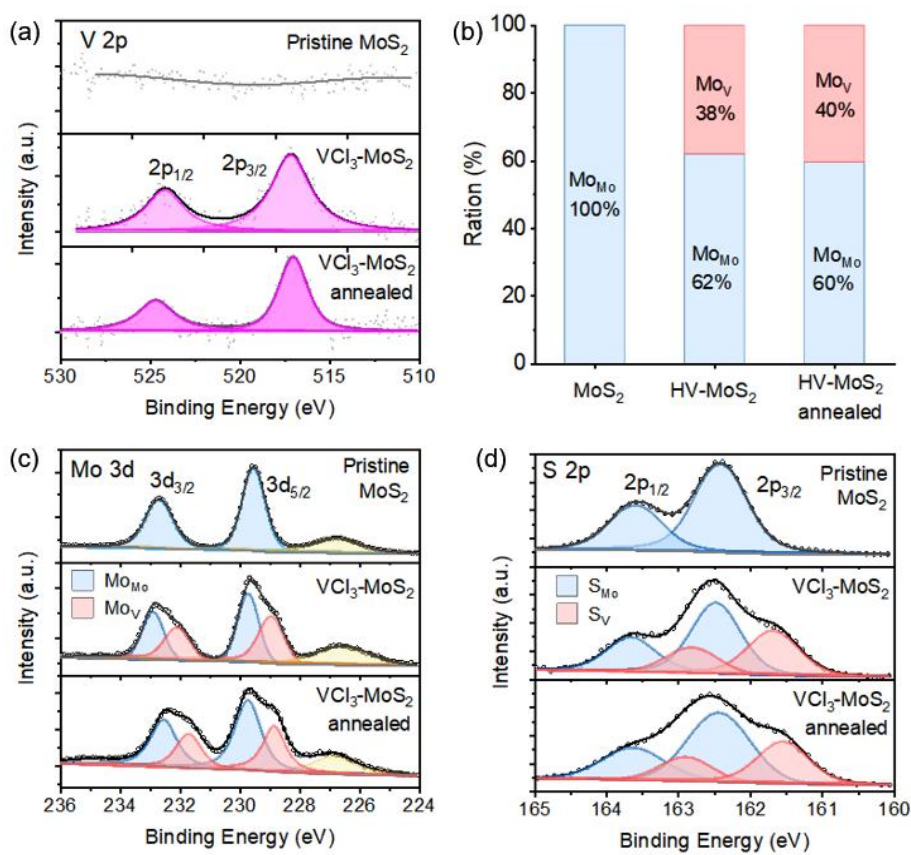

**Figure 2.** XPS analysis of the V-doping effect. (a,c,d) XPS spectra of the V 2p, Mo 3d and S 2p orbits, respectively. (b) The ratio of Mo atoms with different bonding states in the pristine $MoS_2$, $VCl_3$-$MoS_2$ and annealed $VCl_3$-$MoS_2$. $Mo_V$ and $S_V$ represent the Mo and S atoms whose bonding states have been affected by V atoms, and $Mo_{Mo}$ and $S_{Mo}$ represent the unaffected Mo and S atoms.

Raman spectroscopy was used to study how the substituted V atoms affect the structure and electronic properties of $MoS_2$. **Figure 3**a shows the Raman spectra of the pristine $MoS_2$, $V_2O_5$-$MoS_2$, $NH_4VO_3$-$MoS_2$ and $VCl_3$-$MoS_2$. For the lightly doped $V_2O_5$-$MoS_2$, it shows the typical Raman spectrum as the pristine $MoS_2$. Only the characteristic $E_{2g}$ and $A_{1g}$ peaks of $MoS_2$ are observed. However, the Raman spectra of the heavily doped $NH_4VO_3$-$MoS_2$ and $VCl_3$-$MoS_2$ change significantly. First, both the $E_{2g}$ and $A_{1g}$ peaks split into two peaks (379.7 cm$^{-1}$ and 385.3 cm$^{-1}$ for $E_{2g}$, 404.2 cm$^{-1}$ and 409.8 cm$^{-1}$ for $A_{1g}$), and a new peak emerges at 391.8 cm$^{-1}$ (Figure S3a, Supporting Information). These split peaks are assigned to the LO(M) and ZO(M) modes of $MoS_2$, caused by the V atoms induced breakage of lattice symmetry.[26] Second, several new peaks occur in the range of 100 to 300 cm$^{-1}$ (Figure S3b, Supporting Information), caused by the V atoms induced lattice distortion.[27] Third, a new peak appears at 323 cm$^{-1}$, which has not been observed in $MoS_2$ doped by other elements, like Nb[17], Re[13] and Co[23]. We find that the intensity of this new peak is sensitive to the doping concentration of V atoms as it increases with the increasing doping concentration (Figure 3a) and with the adsorption of V species on the flake (Figure S4, Supporting Information). Consequently, this new peak can be identified as the characteristic peak ($C_V$) of the heavily V-doped $MoS_2$.

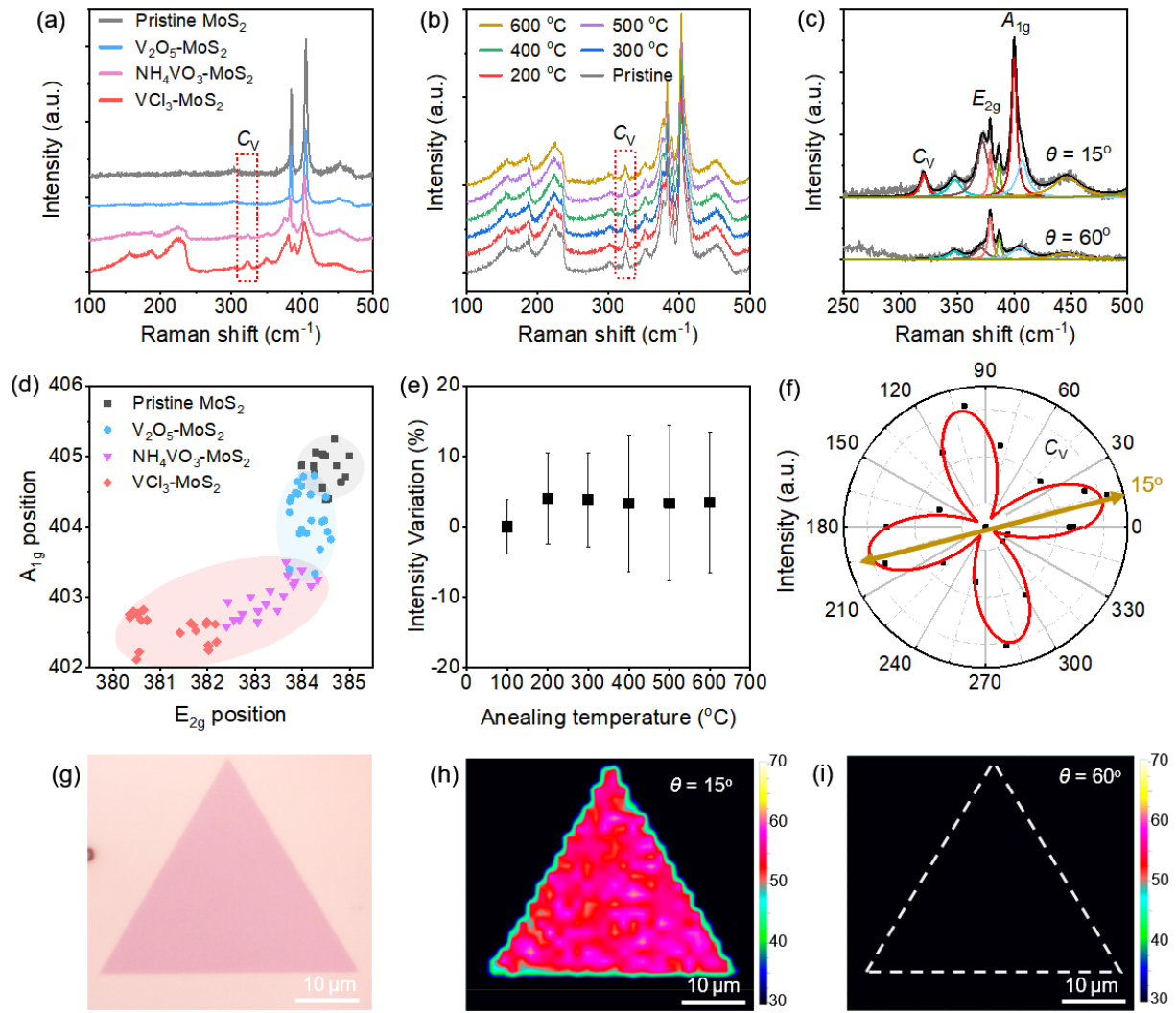

**Figure 3.** Raman spectroscopy of the V-doped MoS$_2$. (a) Raman spectra and (d) statistic data of the $E_{2g}$ and $A_{1g}$ peak positions of the pristine MoS$_2$, V$_2$O$_5$-MoS$_2$, NH$_4$VO$_3$-MoS$_2$ and VCl$_3$-MoS$_2$. (b) Raman spectra of the same VCl$_3$-MoS$_2$ flake and (e) the corresponding intensity variation of the $C_V$ peak after the sample being annealed at different temperatures. (c) Polarized Raman spectra collected when the rotation angle $\theta$ is set at 15° and 60°, and (e) polar plot of the $C_V$ peak intensity as a function of the rotation angle $\theta$. (g) The optical microscope image and (h, i) the polarized Raman mapping of a monolayer VCl$_3$-MoS$_2$ when $\theta$ =15° (h) and $\theta$ = 60° (i).

Furthermore, both the $E_{2g}$ and $A_{1g}$ peaks shift to lower wavenumbers due to the V-doping induced charge transfer and strain in the MoS$_2$ lattice (Figure S3c, Supporting Information). Statistics of the peak positions show that the $A_{1g}$ and $E_{2g}$ peaks of V$_2$O$_5$-MoS$_2$ shift slightly (Figure 3d), indicating that a light p-type doping effect and a small lattice strain were generated in the lightly doped MoS$_2$.[28-30] Nevertheless, the $A_{1g}$ and $E_{2g}$ peaks of NH$_4$VO$_3$-MoS$_2$ shift by 2 cm$^{-1}$ and 1.5 cm$^{-1}$ on average, showing the existence of a heavy p-type doping effect and a large lattice strain. For the most heavily doped VCl$_3$-MoS$_2$, a downshift up to 4 cm$^{-1}$ of the $E_{2g}$ peak is observed, showing the further increased lattice strain as many V atoms are aggregated to form line clusters (Figure S2, Supporting Information). Besides, the VCl$_3$-MoS$_2$ flake kept stable when being annealed at different temperatures ranging from 100 °C to 600 °C (Figure S5, Supporting Information). The Raman spectra as well as the peak positions of the VCl$_3$-MoS$_2$ flake have not changed (Figure 3b), indicating the stability of the V-doping induced lattice distortion, strain and charge transfer. Meanwhile, the intensity of the $C_V$ peak acquired at 323 cm$^{-1}$ maintains constant (Figure 3e), demonstrating that the doping concentration of V atoms kept unchanged at high temperatures. All these Raman results show that the substituted V atoms generate a stable doping effect to modulate the physical property of MoS$_2$.

Next, we investigated the uniformity of V distribution in MoS$_2$. From the HAADF-STEM images, we find that the substituted V atoms are uniformly distributed at the atomic level (Figure 1c-e and Figure S2, Supporting Information). Then, we used the polarized Raman mapping to examine the macroscopic uniformity. The test configuration of the angle-resolved polarized Raman spectroscopy is schematically shown in Figure S6. The intensity of the $C_V$ peak changes with the rotation angle ($\theta$) of the half-wave plate (Figure 3c), and it reaches the maximum when the rotation angle $\theta$ is set at 15º and quenches when $\theta$ is 60º (Figure 3f), demonstrating that the $C_V$ peak intensity is sensitive to the polarization angle of the incident light. Taking the abovementioned doping concentration sensitivity of the $C_V$ peak into account, the uniform intensity profile observed when $\theta$ is 15º (Figure 3h) and the total extinction of the intensity observed when $\theta$ is 60º (Figure 3i) demonstrate that the doping concentration is homogeneous throughout the whole flake. Therefore, the STEM and Raman results confirm the distribution uniformity of substituted V atoms.

We then examined the uniformity of the electronic properties of the heavily doped $VCl_3$-$MoS_2$. A monolayer $MoS_2$ flake with a height of 0.9 nm is shown by the AFM topography image (**Figure 4**a). The whole flake shows a constant surface potential in the scanning Kelvin probe microscopy (SKPM) image (Figure 4b), reflecting that the V-doped $MoS_2$ has a uniform work function.[31,32] Moreover, the whole flake exhibits a homogeneous phase shift in the electrostatic force microscopy (EFM) image (Figure 4c), revealing that the carrier concentration and mobility are uniform throughout the whole flake.[33] Both the SKPM and EFM results confirm that the uniformly distributed V atoms endow the heavily V-doped $MoS_2$ flake with a homogeneous electronic property, making it an appropriate platform for electronic devices.

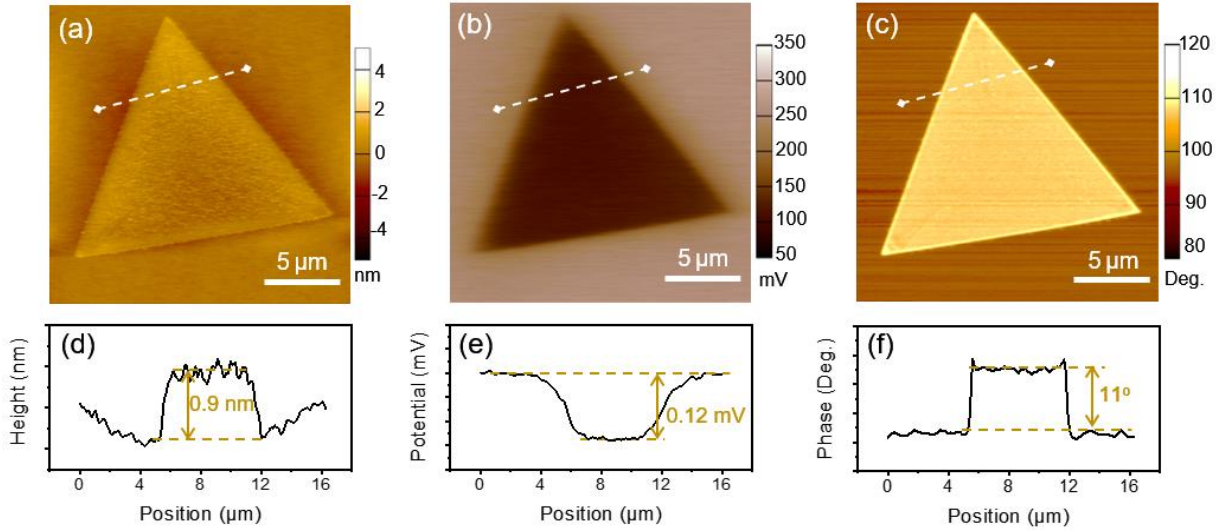

**Figure 4**. (a) AFM topography image, (b) SKPM image, and (c) EFM nap phase image of a monolayer VCl$_3$-MoS$_2$ flake. (d-f) The corresponding line profiles of the height, surface potential and nap phase.

Finally, we explored the device applications of V-doped MoS$_2$ by fabricating field-effect transistors (FETs) using the heavily doped VCl$_3$-MoS$_2$ as the channel material (**Figure 5**a). Previous results show that V atoms can induce energy states deep inside the bandgap of MoS$_2$[34], so V atoms may act as trap centers to trap/release carriers. When a positive gate voltage pulse is applied to the device, the whole channel is filled with free electrons, and these free electrons would be easily caught by the trap centers (Figure 5b). Then the concentration of free electrons decreases significantly after the positive pulse. On the contrary, the channel is fully depleted when a negative pulse is applied. The trapped electrons will be released into the channel to increase the concentration of free electrons after the negative voltage pulse. Therefore, with abundant V atoms acting as carrier trap centers in the heavily V-doped MoS$_2$, the channel conductance of the transistor can be regulated by the gate voltage pulses as the trapped electrons do not participate in conduction.[35] As a result, the potentiation and depression behaviors of synapses are mimicked by the transistor made of the heavily V-doped MoS$_2$. To verify this carrier-trapping-detrapping mechanism, electron concentrations in the lightly doped V$_2$O$_5$-MoS$_2$ and heavily doped VCl$_3$-MoS$_2$ are calculated from the transfer curves (Figure S7, Supporting Information). The electron concentration decreases by more than 10 times in VCl$_3$-MoS$_2$ because many electrons are trapped by V atoms. Besides, during a continuous sweeping of the bias voltage from -80 V → 80 V → -80 V, hysteresis loops are observed only in the transistor made of VCl$_3$-MoS$_2$, reflecting that nonvolatile change of the channel conductance can only be realized in the heavily V-doped MoS$_2$. These results show that the high doping concentration is essential to implement synaptic transistors when using V-doped MoS$_2$ as the channel material.

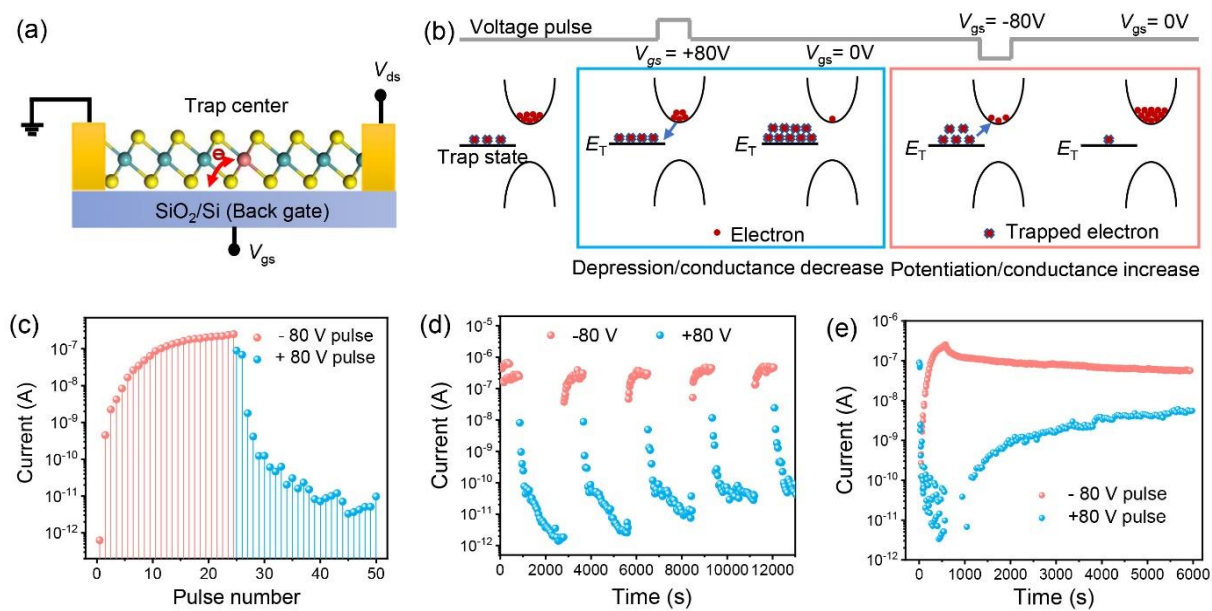

**Figure 5.** Synaptic transistor based on the heavily doped VCl$_3$-MoS$_2$. (a) Schematic of the gate-tunable synaptic transistor. (b) The carrier trapping and detrapping process stimulated by the positive and negative pulses. (c) The synaptic potentiation (the 1$^{st}$ to the 25$^{th}$ pulse) and depression (the 26$^{th}$ to the 50$^{th}$) processes stimulated by the -80 V and 80 V voltage pulses. Note that the current was measured after the gate pulses ($V_g$ = 0 V). (d) The repetition of the potentiation and depression processes with alternant negative and positive pulses. (e) The retention curve of the high-conductance and low-conductance states.

We then investigated the synaptic behavior of the transistor made of the heavily V-doped MoS$_2$. The current in the transistor continuously increases over 10$^5$ times from 6.3×10$^{-13}$ A to 2.5×10$^{-7}$ A when 25 rounds of negative gate voltage pulses (-80 V) are applied (Figure 5c). Then the current gradually decreases back to about 5×10$^{-12}$ A when another 25 rounds of positive gate voltage pulses (+80 V) are applied. Hence, both the synaptic potentiation and depression responses are simulated by the transistor made of heavily V-doped MoS$_2$. The ratio between the high conductance state and the low conductance state is above 10$^4$ after 5 cycles, and the similar curve shape shows the good repeatability of this synaptic transistor (Figure 5d). Besides, the high-conductance and low-conductance states can be easily distinguished after 6000 s (Figure 5e). All these results show that the synaptic transistor has a good endurance and retention performance, and it can mimic the repetitive learning process of synapses, proving that heavily V-doped MoS$_2$ is a unique platform to construct artificial synaptic devices.

## 3. Conclusion

In summary, we have developed an effective method to realize the controlled V-doping of MoS$_2$. The doping concentrations of V atoms are tunable in a wide range of 0.3-13.1 at%. The substituted V atoms are uniformly distributed in the V-doped MoS$_2$ to create a stable and homogeneous doping effect and electronic property, and they can serve as the trap centers to modify the electrical conductivity of V-doped MoS$_2$. For the first time, gate-tunable synaptic transistors were fabricated by using the heavily doped VCl$_3$-MoS$_2$ as the channel material, which well simulates the synaptic potentiation, depression and repetitive learning processes of a synapse. This work paves a way to realize the controlled doping of 2D materials and offers a unique platform to design functional electronic devices like artificial synaptic transistors.

## 4. Experimental Section

*Growth of monolayer V-doped MoS$_2$*: A tube furnace with two heating zones was used for the CVD growth of MoS$_2$. Sulfur powder (150 mg, 99.5%, Sigma-Aldrich) loaded in a quartz boat was placed upstream in the first zone and heated to 220 °C during the growth process. Another quartz boat with MoO$_3$ powder (3 mg, 99.9%, Sigma-Aldrich) inside was placed at the center of the second heating zone. The Si substrate with 285 nm thick SiO$_2$ was put just above the MoO$_3$ powder and faced down. V sources, including V$_2$O$_5$, NH$_4$VO$_3$ and VCl$_3$ powders, were placed in the middle between S and MoO$_3$, about 6 cm away from the growth substrate. The tube was flushed with Ar at a rate of 1000 sccm for 10 min, and then the rate was set at 80 sccm during the following growth process. The second zone was heated to 700 °C at a rate of 50 °C /min and maintained for 10 min for the growth of V-doped MoS$_2$.

*Characterization*: Optical microscope pictures were taken by an optical microscopy (Carl Zeiss Microscopy, USA). Raman and PL spectra were collected by a Raman spectroscopy (Horiba LabRAB HR Evolution, Japan), using a 532 nm laser excitation with a spot size of ~1 μm. For the angle-resolved polarized Raman measurements, a parallel polarization test configuration was adopted.[36] Chemical elemental analyses of the samples were conducted by XPS (monochromatic Al Kα X-rays, PHI VersaProbe II, Japan). HAADF-STEM images were taken by an aberration-corrected TEM (FEI Titan Cube Themis G2, USA) operated at 60 kV with a resolution of about 1 Å. EFM and SKPM were performed by AFM (Cypher ES, Oxford Instruments, UK) using a conductive tip.

*Device fabrication and measurements*: FET devices were fabricated using a laser writing system (miDALIX, DaLI, German). In brief, a drop of AZ5214 photoresistor (PR) was spin-coated onto the $SiO_2$/Si substrate with the samples attached (2000 rpm for 1 min) and baked at 125 °C for 1 min. Photo-lithography was then performed using PR as the positive resist, followed by the metal deposition and lift-off processes. Metal electrodes were made of 5 nm Pd and 50 nm Au, which were deposited using an e-beam evaporation system. The device measurements were performed using a semiconductor characterization system (4200-SCS, Keithley, USA) in a vacuum probe station ($10^{-5}$ mBar, Lakeshore, USA).

**Acknowledgments**

The authors acknowledge the supports by the National Natural Science Foundation of China (Grant Nos. 51722206, 51920105002, 51991340, 11974156 and 51991343), the Youth 1000-Talent Program of China, Guangdong Innovative and Entrepreneurial Research Team Program (Program No. 2017ZT07C341 and No. 2019ZT08C044), the Bureau of Industry and Information Technology of Shenzhen for the "2017 Graphene Manufacturing Innovation

Center Project" (Project No. 201901171523), and the Shenzhen Basic Research Project (No. JCYJ20200109144620815 and KQTD20190929173815000). This work was also assisted by SUSTech Core Research Facilities, especially technical support from Pico-Centre that receives support from Presidential fund and Development and Reform Commission of Shenzhen Municipality.

**Author contributions**

Jingyun Zou, Zhengyang Cai, and Yongjue Lai contributed equally to this work.

**Supporting Information**

Supporting Information is available from the Wiley Online Library or from the author.

**Competing interests:** The authors declare no competing interests.